\documentclass[aip,jcp,preprint,floatfix,a4paper]{revtex4-1}

\usepackage{graphicx}
\usepackage[sort&compress]{natbib}
\usepackage[sumlimits]{amsmath}
\usepackage{amsfonts}
\usepackage{amssymb}
\usepackage{dcolumn}
\usepackage{subdepth}
\usepackage{color}
\usepackage{enumitem}
\usepackage{hyperref}
\usepackage{lipsum}
\usepackage{subdepth}
\usepackage{acro}
\usepackage{stackrel}
\usepackage{multirow}
\usepackage{float}

\DeclareAcronym{QM}{
  short = QM ,
  long  = quantum-mechanical,
  class = abbrev
}
\DeclareAcronym{TCF}{
  short = TCF ,
  long  = time correlation function,
  class = abbrev
}
\DeclareAcronym{RIXS}{
  short = RIXS ,
  long  = resonant inelastic X-ray scattering,
  class = abbrev
}

\DeclareAcronym{XAS}{
  short = XAS ,
  long  = X-ray absorption spectrum,
  class = abbrev
}

\DeclareAcronym{DOF}{
  short = DOF ,
  long  = degree of freedom,
  long-plural-form = degrees of freedom,
  class = abbrev
}

\DeclareAcronym{IR}{
  short = IR ,
  long  = infra-red,
  class = abbrev
}

\DeclareAcronym{PI}{
  short = PI,
  long  = path integral,
  class = abbrev
}

\DeclareAcronym{QCF}{
  short = SCF,
  long  = shift correction factor,
  class = abbrev
}

\DeclareAcronym{BOA}{
  short = BOA,
  long  = Born-Oppenheimer approximation,
  class = abbrev
}

\DeclareAcronym{MD}{
  short = MD,
  long  = molecular dynamics,
  class = abbrev
}

\DeclareAcronym{CMD}{
  short = CMD,
  long  = centroid molecular dynamics,
  class = abbrev
}

\DeclareAcronym{RPMD}{
  short = RPMD,
  long  = ring polymer molecular dynamics,
  class = abbrev
}

\DeclareAcronym{TRPMD}{
  short = TRPMD,
  long  = thermostatted RPMD,
  class = abbrev
}

\DeclareAcronym{PES}{
  short = PES,
  long  = potential energy surface,
  class = abbrev
}

\DeclareAcronym{NRPMD}{
  short = NRPMD,
  long  = nonadiabatic ring polymer molecular dynamics,
  class = abbrev
}

\DeclareAcronym{DCL}{
  short = DCL,
  long  = dynamical classical limit,
  class = abbrev
}

\DeclareAcronym{ACL}{
  short = ACL,
  long  = averaged classical limit,
  class = abbrev
}

\DeclareAcronym{SCL}{
  short = SCL,
  long  = statical classical limit,
  class = abbrev
}

\DeclareAcronym{EOM}{
  short = EOM,
  long  = equation of motion,
  long-plural-form = equations of motion,
  class = abbrev
}

\DeclareAcronym{MM}{
  short = MM,
  long  = molecular mechanics,
  class = abbrev
}

\newcommand{\QM}{\ac{QM}}
\newcommand{\TCF}{\ac{TCF}}
\newcommand{\TCFs}{\acp{TCF}}

\newcommand{\DOFs}{\acp{DOF}}
\newcommand{\IR}{\ac{IR}}
\newcommand{\PI}{\ac{PI}}

\newcommand{\BOA}{\ac{BOA}}
\newcommand{\RPMD}{\ac{RPMD}}
\newcommand{\TRPMD}{\ac{TRPMD}}

\newcommand{\CMD}{\ac{CMD}}

\newcommand{\PES}{\ac{PES}}
\newcommand{\PESs}{\acp{PES}}
\newcommand{\NRPMD}{\ac{NRPMD}}
\newcommand{\DCL}{\ac{DCL}}
\newcommand{\ACL}{\ac{ACL}}

\newcommand{\EOMs}{\acp{EOM}}

\newcommand{\mOp}[1]{\hat{#1}}
\newcommand{\mVec}[1]{\boldsymbol{\mathrm{#1}}}
\newcommand{\mBra}[1]{\langle #1 |}
\newcommand{\mKet}[1]{| #1 \rangle}

\newcommand{\Eq}[1]{Eq.\,(\ref{#1})}
\newcommand{\Eqs}[1]{Eqs.\,({#1})}
\newcommand{\Sec}[1]{Sec.\,\ref{#1}}
\newcommand{\mMat}[1]{{\boldsymbol{\mathbb{#1}}}}
\newcommand{\FT}[1]{S}
\newcommand{\e}{\mathrm{e}}
\newcommand{\im}{\mathrm{i}}
\newcommand{\diff}{\mathrm{d}}
\newcommand{\Fig}[1]{Fig.\,\ref{#1}}

\renewcommand{\Re}{\mathrm{Re}}
\newcommand{\Tr}{\mathrm{Tr}}

\renewcommand{\l}{\lambda}
\newcommand{\w}{\omega}
\newcommand{\W}{\Omega}
\newcommand{\T}[1]{#1^{\mathrm{T}}}

\newcommand{\nn}{\nonumber}
\renewcommand{\H}{\mathcal{H}}
\newcommand{\x}{\mVec{x}}
\newcommand{\p}{\mVec{p}}
\newcommand{\D}{\mVec{\Delta}}

\newcommand{\Q}{\mVec{Q}}
\renewcommand{\P}{\mVec{\Pi}}

\newcommand{\MT}{\mMat{T}}
\newcommand{\MW}{\mMat{W}}
\renewcommand{\t}[1]{\tilde{#1}}
\newcommand{\grad}{\mVec{\nabla}}
\newcommand{\U}{\mathcal{U}}

\begin{document}

%SK: Oliver doesn't like the beyond because all other presented methods do not really go beyond the Matsubara dynamics. They are approximations to it and the modified method is ad hoc. He suggests:
%SI: Never saw a title constructed out of two sentences
\title{Simulating vibronic spectra via Matsubara dynamics: coping with the sign problem}

\affiliation{Institute of Physics, Rostock University, Albert-Einstein-Str. 23-24, 18059 Rostock, Germany}
\author{Sven Karsten}
\author{Sergei D. Ivanov}
\email{sergei.ivanov@uni-rostock.de}
\author{Sergey I.\ Bokarev}
\author{Oliver K\"uhn}
\date{\today}                                           

\begin{abstract}

Measuring the vibronic spectrum probes dynamical processes in molecular systems.
When interpreted via suitable theoretical tools, the experimental data provides comprehensive information about the system in question.
For complex many-body problems, such an approach usually requires the formulation of proper classical-like approximations, which is particularly challenging if multiple electronic states are involved.
In this manuscript, we express the imaginary-time shifted \TCF\ and, thus, the vibronic spectrum in terms of the so-called Matsubara dynamics, which combines quantum statistics and classical-like dynamics.
In contrast to the existing literature, we invoke a local harmonic approximation to the potential allowing an analytical evaluation of integrals.
By subsequently applying the Matsubara approximation, we derive a generalization of the existing Matsubara method to multiple \PESs, which, however, suffers from the sign problem as its single-\PES\ counterpart does.
The mathematical analysis for two shifted harmonic oscillators suggests a new modified method to simulate the standard correlation function via classical-like dynamics.
Importantly, this modified method samples the thermal Wigner function without suffering from the sign problem and it yields an accurate approximation to the vibronic absorption spectrum, not only for the harmonic system, but also for an anharmonic one.
\end{abstract}

\maketitle

%---------------------------------
\section{Introduction}
\label{sec:Intro}
%---------------------------------

Spectroscopic methods constitute the cornerstone of experimental molecular physics and physical chemistry.~\cite{Mukamel-Book,Kuehn-Book,Hamm-Zanni-Book}
Constant sharpening of the energy and time resolution provides increasingly detailed insight into photophysical and photochemical processes.
Complementing measurements by proper theoretical tools enables understanding of the underlying microscopic phenomena. 
Simple models,~\cite{Kuehn-Book, Mukamel-Book} on the one hand, often fail to capture essential features, whereas solving the time-dependent Sch\"odinger equation numerically exactly,~\cite{MCTDH-Book} on the other hand, is not feasible for large systems due to the infamous curse of dimensionality.
Therefore, compromises between the both extremes are to be found.
A remarkable pathway is to approximate the exact quantum dynamics via (quasi-)classical methods motivated by the provided intuitive access into the atomistic picture.
For instance, imaginary-time \PI\ techniques have proven themselves to rigorously account for static quantum effects
profiting from the classical-like \EOMs.~\cite{Feynman-Book-1965,Schulman-Book,Marx-Book,Tuckerman-Book}
These approaches are successfully employed for electronic ground-state properties and purely vibrational spectra.~\cite{Ivanov-NatChem-2010,Witt-PRL-2013,Olsson-JACS-2004,Gao-AR-2002}
Here, the quantum time evolution has a direct classical analogue since only a single electronic \PES\ is of relevance.
In contrast, vibronic spectra are intrinsically more quantum due to electronic \DOFs\ undergoing transitions between discrete levels.
Thus in a simulation, at least two \PESs\ need to be explicitly accounted for; such a situation has no classical analogue.

A number of trajectory-based methods treating dynamics on several \PESs\ has been suggested (see, e.g., Refs.~\onlinecite{Stock_Thoss-2005,Tully-JCP-2012,Tavernelli-ACR-2015} for review), mostly addressing the rates of non-adiabatic transitions, while leaving the problem of vibronic spectra aside.
Further, \NRPMD\ techniques have been developed,~\cite{Richardson-JCP-2013,Ananth-JCP-2013} which are based on the mapping approach introduced by Stock and Thoss.\cite{Thoss-PRA-1999,Stock_Thoss-2005,Ananth-JCP-2010}
Although these methods are suitable for vibronic spectra, given an efficient simulation protocol is provided,~\cite{Richardson-CP-2016} such an application has not been commonly discussed.
Similarly, \PI\ approaches for many \PESs\ without mapping variables~\cite{Schwieters_JCP_1999,Alexander_CPL_2001,Schmidt_JCP_2007,Shushkov_JCP_2012,Lu_JCP_2017} have addressed mostly non-adiabatic effects on static properties or reaction rates but not vibronic spectra.

In a recent publication, Ref.~\citenum{Karsten_JCP_2018}, we have suggested a generalized formalism to address the vibronic spectra of complex systems.
Within this approach, many commonly known correlation functions such as the Kubo~\cite{Kubo1957} or Schofield~\cite{Schofield-PRL-1960} \TCFs\ as well as the more pragmatic \DCL\ one~\cite{Rabani-JCP-1998,Egorov-JCP-1998,Karsten-JPCL-2017,Karsten-JCP-2017} are naturally recovered. 
Additionally, this formulation enables constructing completely new \TCFs\ and it has been demonstrated for model systems that some newly suggested \TCFs\ can yield numerical protocols that outperform the well-established ones.
This has led to the conclusion that there is no unambiguously favorable
\TCF\ when it comes to a practical consideration of the vibronic spectrum, which is in contrast to \IR\ spectroscopy, where the 
Kubo \TCF\ is to be preferred.~\cite{Ramirez2004,Craig-JCP-2004}
In Ref.~\citenum{Karsten_JCP_2018}, an adiabatic quasi-classical approximation to the imaginary-time shifted \TCF, which is the main ingredient of the generalized formalism, has been carried out.
In the spirit of the standard \RPMD\ method,~\cite{Craig-JCP-2004} a numerically exact expression for the \TCF\ at time zero has been derived.
The subsequent dynamics has been constructed to conserve the quantum Boltzmann density, but it suffers from infamous artificial harmonic spring oscillations, intrinsic to quasi-classical \PI\ techniques.~\cite{Witt-JCP-2009}

During the last years, several attempts have been performed to develop improved approximations to vibrational quantum dynamics based on the so-called Matsubara dynamics.~\cite{Hele-JCP-2015,Hele-JCP-2015_2,Hele-MP-2016}
This methodology starts from the observation that only smooth imaginary-time paths contribute to canonical thermal averages, whereas jagged or discontinuous paths are sufficiently suppressed by the Boltzmann operator.
Assuming that the path remains smooth even if it undergoes dynamics, i.e.\ the Matsubara approximation, one can rigorously derive classical-like \EOMs\ leaving the initial density unchanged.
Although the Matsubara dynamics ansatz yields a reasonable approximation to the Kubo \TCF\ for small systems, it is not yet applicable to high-dimensional problems due to the infamous sign problem.
Still, it has been shown that popular methods such as \RPMD, \TRPMD~\cite{Rossi-JCP-2014} as well as \CMD~\cite{Cao-JCP-1993} can be viewed as feasible approximations to the Matsubara dynamics.
Thus, it is natural to expect that other even more powerful approximations to nuclear quantum dynamics can be derived on its basis, making this approach particularly promising.

However, so far the Matsubara dynamics method has not been employed for vibronic spectra.
In this manuscript, we present how this spectroscopic observable can be approximately evaluated making use of the Matsubara dynamics.
First, the exact imaginary-time shifted \TCF\ is rigorously reformulated in terms of a phase-space integration, see \Sec{sec:phase_space}.
Second, alternatively to existing literature, a local harmonic approximation to the potential is involved in \Sec{sec:time-ind-part}.
Together with the Matsubara approximation itself, this leads to a generalization of the established Matsubara method to multiple \PESs.
Third, classical-like \EOMs\ are derived in \Sec{sec:time-dep-part}, leaving a flexibility for a choice of Hamilton function generating the dynamics.
Two reasonable variants are motivated in \Sec{sec:choosing}, followed by a discussion of common approaches to circumvent the sign problem in \Sec{sec:sign-problem}.
Finally, favorable simulation scenarios are suggested, based on the mathematical analysis for an analytically solvable system consisting of two shifted harmonic oscillators in \Sec{sec:Results}.
Moreover, a modified method to simulate the standard correlation function without suffering from the sign problem is outlined.
Importantly, this method samples directly the thermal Wigner function of the harmonic oscillator.
All considered approaches are subsequently applied to a one-dimensional model system at ambient temperature, see \Sec{sec:harmonic_num}.
In \Sec{sec:modified_gen}, the modified method is generalized to anharmonic systems and is demonstrated to sample accurately the thermal Wigner function for a quartic expansion of the Morse potential.
Interestingly, although designed for the harmonic case, the modified method yields vibronic spectra very similar to those obtained with the Matsubara dynamics method for the general case, see \Sec{sec:Morse_Spec}.
In \Sec{sec:Conclusions}, the manuscript is concluded and some perspectives for future research are given.

%---------------------------------
\section{Theory}
\label{sec:Theory}
%---------------------------------

Traditionally, the heart of (N)RPMD methodologies is the Kubo-transformed \TCF, however, we have shown recently that when vibronic spectra are addressed, the choice of the correlation function is not as unambiguous as it is for vibrational dynamics.~\cite{Karsten_JCP_2018}
The cornerstone of the generalized \TCF\ suggested therein is the formally exact imaginary-time shifted \TCF, $C_\l(t)$ (see \Eq{eq:shifted_TCF}).
Here, we generalize the well-established Matsubara dynamics~\cite{Hele-JCP-2015} to vibronic spectroscopy and derive it from $C_\l(t)$ by means of a local harmonic approximation to the potential.

%----------------------------------
\subsection{Setting the stage}
\label{sec:phase_space}
%----------------------------------

\begin{figure}
\includegraphics[width=0.6\columnwidth]{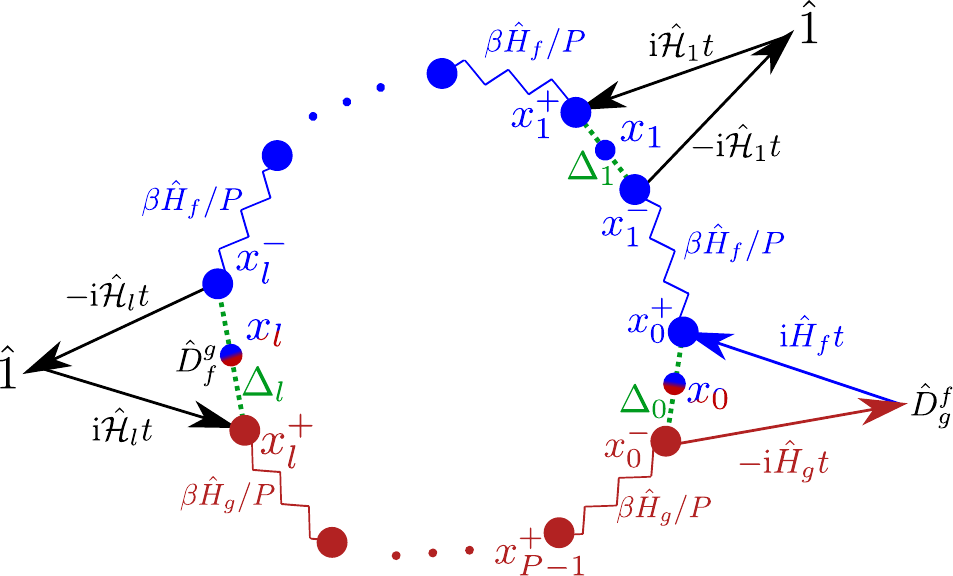}
\caption{\label{fig:star}
Visualization of the structure of the correlation function in \Eqs{\ref{eq:phase-space}-\ref{eq:time-dep-part}}.
The jagged lines and straight arrows represent imaginary- and real-time propagations with length $\beta/P$ and $t$, respectively.
The red and the blue color indicates initial and final electronic state, correspondingly.
}
\end{figure}

The starting point is the imaginary-time shifted correlation function for a particular $\l \in [0,\beta]$~\cite{Karsten_JCP_2018}
 \begin{align}
 \label{eq:shifted_TCF}
 C_{\l}(t)=\frac{1}{Z}  \Tr \left [  \e^{-(\beta-\l) \mOp{H}_g}\mOp{D}^{g}_f \e^{-\l\mOp{H}_f} \e^{\im \mOp{H}_f t / \hbar} \mOp{D}^{f}_g  \e^{-\im \mOp{H}_g t/\hbar}  \right  ]   \enspace,
 \end{align}
where the \BOA\ is assumed, $\Tr [\bullet]$ stands for a trace in the nuclear Hilbert space, $\mOp{H}_a$ corresponds to the nuclear Hamiltonian with the \PES\ $\mOp{V}_a$ of the $a$-th adiabatic state, where $a=g,f$ and $\mOp{D}^{g}_f \equiv (\mOp{D}^{f}_g)^*$ is the transition dipole moment.
Note that for $\l=0$, the standard correlation function $C_0(t)$ is recovered, whose Fourier transform directly yields the absorption spectrum.
For the sake of brevity, we restrict ourselves to a 1D system, described by position and momentum operators $\mOp{x}$ and $\mOp{p}$ in nuclear space and two discrete electronic states $g$ and $f$; the generalization to a many-dimensional multi-level system is straightforward.
To start, the shift in imaginary time is equidistantly discretized, i.e.\ $\l=l \beta/P$, where $P>0$ is a natural number and $\l$ and $l$ are used synonymously throughout the manuscript.
The nuclear trace in \Eq{eq:shifted_TCF} is performed in the eigenstate basis of $\mOp{x}$ over coordinates $x_0^{-}$ and an additional spatial closure, i.e.\ an integral over $x_{0}^{+}$, is inserted. 
Subsequently, a canonical variable substitution to midpoint and difference variables is performed, i.e.\ $x_0:=(x_{0}^{+}+x_{0}^{-})/2$ and $\Delta_0:=x_{0}^{+}-x_{0}^{-}$, respectively, see \Fig{fig:star} for a sketch.
The resulting expression reads
\begin{align}
 C_{\l}(t)  & = \frac{1}{Z} \int \diff x_0 \int  \diff \Delta_0    \mBra{x_{0}^{-}} \e^{-(P-l)\beta\mOp{H}_g/P}\mOp{D}^{g}_f \e^{-l\beta\mOp{H}_f/P} \mKet{x_{0}^{+}}  \mBra{x_{0}^{+}} \e^{\im \mOp{H}_f t / \hbar} \mOp{D}^{f}_g  \e^{-\im \mOp{H}_g t/\hbar} \mKet{x_{0}^{-}}   \enspace.
\end{align}
Following the standard imaginary-time \PI\ approach,~\cite{Chandler-JCP-1981,Tuckerman-Book} each exponential term $\exp(-j\beta\mOp{H}_a/P)$, with $j=P-l,l$ can be expressed as a product of $j$ identical factors $\exp(-\beta\mOp{H}_a/P)$.
Further, unity operators in the form
\begin{align}
\label{eq:Unity}
 \mOp{1}  & =  \int \diff x_i \int \diff \Delta_i   \mKet{x_{i}^{+}} \mBra{x_{i}^{+}} \e^{\im \mOp{\H}_i t / \hbar}    \e^{-\im \mOp{\H}_i t/\hbar} \mKet{x_{i}^{-}} \mBra{x_{i}^{-}} 
\end{align}
are inserted in between those factors, where the coordinates $x_i^{\pm}$ correspond to the sum and difference variables $x_i$ and $\Delta_i$, respectively, as it is visualized in \Fig{fig:star}.
The $\mOp{\H}_i$ are, in principle, arbitrary Hermitian operators, $i=1,\ldots,P-1$.
However, for the present purpose they are limited to a non-relativistic Hamiltonian form $\mOp{\H}_i=\mOp{p}^2/2m+\mOp{\mathcal{V}}_i(\mOp{x})$, where  $m$ is the nuclear mass and particular choices for the potentials $\mOp{\mathcal{V}}_i$ are suggested in \Sec{sec:choosing}.
To obtain the structure of the Wigner transform,~\cite{Wigner-PR-1932,DeAlmeida-PR-1998} which will be of use later, scalar unities are inserted
\begin{align}
 1& = \intop \diff {\Delta}'_i \, \delta({\Delta}'_i - {\Delta}_i) = \frac{1}{2\pi \hbar} \int \diff {\Delta}'_i \intop \diff {p}_i \, \e^{\im p_i(\Delta'_i-\Delta_i)/\hbar}
\end{align}
for each $i$.
Finally, time-(in)dependent parts can be singled out, leading to the compact form of a classical-like phase-space integral
\begin{align}
\label{eq:phase-space}
 C&_{\l}(t)= \int \diff \x \int \diff \p \, A_l(\x,\p) B_l(\x,\p,t)
 \enspace.
\end{align}
Here, the time-independent part reads
\begin{align}
\label{eq:time-ind-part}
A &_l(\x,\p) :=  \nn \\
&\frac{1}{Z (2 \pi \hbar)^{P}}    \int \diff \D  \,  \e^{\im \T{\p}\D/\hbar} \frac{1}{2}[{D}^{g}_f (x_{l}^{+})+{D}^{g}_f (x_{l}^{-})]   \prod_{i=l}^{P-1} \mBra{x_{i+1}^{-}} \e^{-\beta\mOp{H}_g/P} \mKet{x_{i}^{+}}   \prod_{i=0}^{l-1} \mBra{x_{i+1}^{-}} \e^{-\beta\mOp{H}_f/P} \mKet{x_{i}^{+}}
\enskip ,
\end{align} 
where ${\x}:=\T{(x_0,\ldots,x_{P-1})}$ (likewise for $\D$ and $\p$) and all the vectors obey the cyclic condition $P \mapsto 0$.
Note that the dipole is evaluated at both $x_{l}^{\pm}$ since the unity with the index $l$, \Eq{eq:Unity}, can be inserted on the both sides of the dipole operator.
Having \Fig{fig:star} at hand, it becomes clear that the midpoint coordinates $x_i$ form a continuous path in the limit $P\to\infty$.
Analogously to common imaginary-time \PI\ approaches, these midpoints will be referred to as beads in the following. 

The time-dependent part takes the form
\begin{align}
\label{eq:time-dep-part}
B&_l(\x,\p,t)  :=  \int \diff \D  \,  \e^{-\im \T{\p}\D/\hbar}
 \prod_{i=0}^{P-1}  \mBra{x_{i}^{+}} \e^{\im \mOp{\H}_i t / \hbar}   \mOp{\mathcal{O}}_i  \e^{-\im \mOp{\H}'_i t/\hbar} \mKet{x_{i}^{-}} 
\enspace ,
\end{align}
where $\mOp{\mathcal{O}}_i=\mOp{1}, \forall i>0$ and $\mOp{\mathcal{O}}_0:=\mOp{D}^{f}_g$.
Importantly, for $i=0$ the Hamiltonians are fixed $\mOp{\H}'_0:=\mOp{H}_g$ and $\mOp{\H}_0:=\mOp{H}_f$, whereas the
$\mOp{\H}'_i \equiv \mOp{\H}_i, \forall i>0$ remain arbitrary.
Note further that at time zero the expression collapses to $B_l(\x,\p,0) \equiv  {D}^{f}_g(x_0)$ as it can be easily proven.
The \Eqs{\ref{eq:phase-space}-\ref{eq:time-dep-part}} remain exact and possible approximations to them are discussed in \Sec{sec:time-ind-part} and \ref{sec:time-dep-part}.

\subsection{The time-independent part}
\label{sec:time-ind-part}

\begin{figure}
\includegraphics[width=0.8\columnwidth]{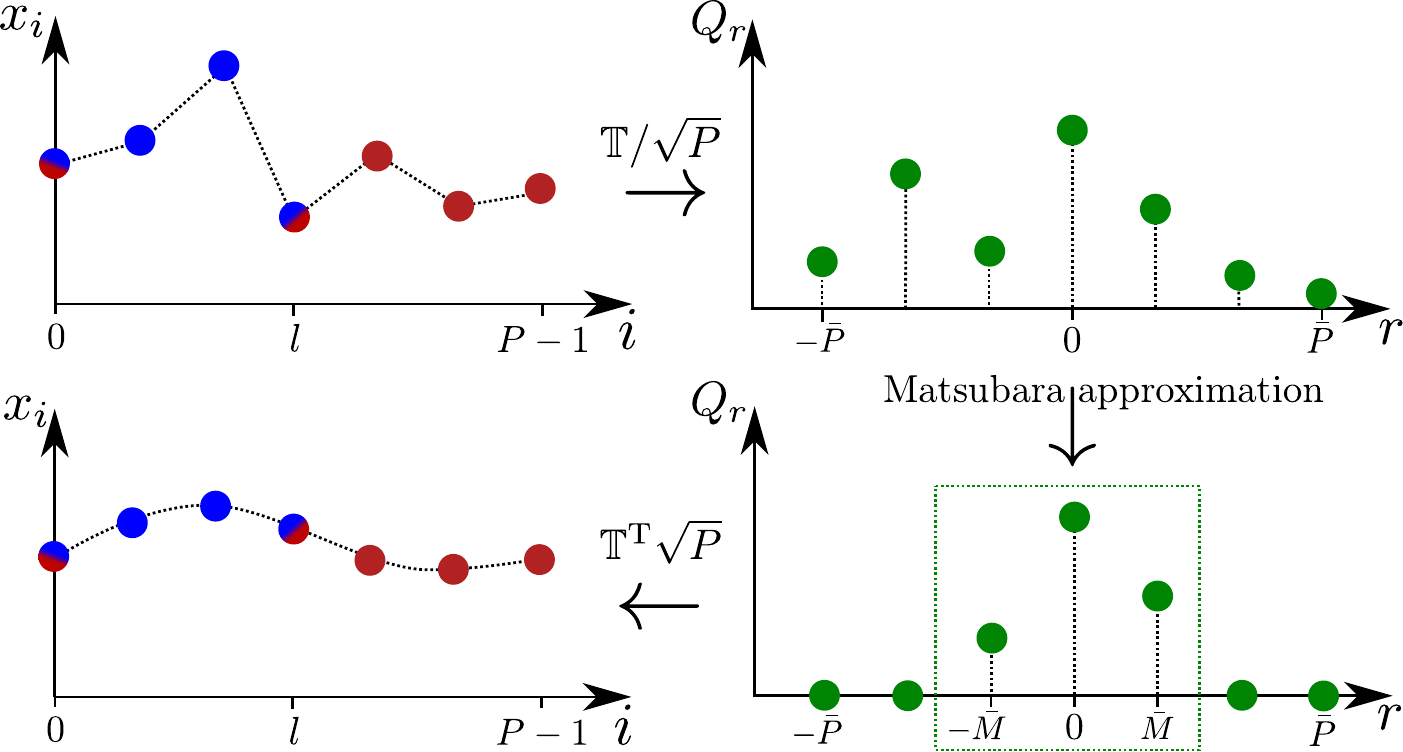}
\caption{\label{fig:path_NM_MM} A sketch illustrating how the Matsubara approximation affects an imaginary-time path.
An arbitrary jagged imaginary-time path (upper left) is transformed to the normal modes' coordinates via the matrix $\MT$ (upper right). After restricting to the $M$ lowest modes (bottom right), i.e.\ the Matsubara approximation, one obtains a smooth path via the back transform (bottom left).
Note that the normal mode index $r$ runs over $-\bar{P}, \ldots, \bar{P}$, where, similarly to $\bar{M}$, $\bar{P}:=(P-1)/2$.
}
\end{figure}

First, each density matrix element in the time-independent part is evaluated via the symmetric Trotter factorization in the usual imaginary-time \PI\ fashion, yielding
\begin{align}
\label{eq:Trotter}
& \mBra{x_{i+1}^{-}} \e^{-\beta\mOp{H}_a/P} \mKet{x_{i}^{+}}  =   \lim_{P \to \infty} \left ( \frac{ \beta m \W^2_P}{2 \pi} \right )^{1/2}   \exp\left\{ -\frac{\beta}{2P} \left [ V_a(x_{i+1}^{-})+V_a(x_{i}^{+})  \right ] - \beta \frac{m \W^2_P}{2} \left [ x_{i+1}^{-}-x_{i}^{+}  \right ]^2\right\}
\enskip, 
\end{align}
where $\W_P:=\sqrt{P}/\beta \hbar$ is the standard ring-polymer chain frequency.
Now the {\it first} approximation is employed, namely that the potential is assumed to be {\it locally} harmonic
\begin{align}
\label{eq:V_harm}
V_a(x_{i}^{\pm}) \approx V_a(x_{i}) \pm \frac{\partial V_a}{\partial x_i} \frac{\Delta_i}{2} +\frac{1}{2}  \frac{\partial^2 V_a}{\partial x_i^2} \left ( \frac{\Delta_i}{2}\right )^2
\end{align}
and the transition dipole to be locally linear
\begin{align}
\label{eq:dip_lin}
\frac{1}{2}[{D}^{g}_f (x_{l}^{+})+{D}^{g}_f (x_{l}^{-})] \approx {D}^{g}_f (x_{l})
\enskip.
\end{align} 
These assumptions enable the analytical integration over the difference variables, see Supplement for a detailed derivation.
%
%TODO: Check the section number
It is useful to employ the normal mode coordinates of the free-particle ring polymer,~\cite{Tuckerman-Book,Witt-JCP-2009} $\Q:=\MT \x /\sqrt{P}$ and $\P:=\MT \p /\sqrt{P}$, where the orthogonal transformation matrix $\MT$ represents the discrete Fourier transform with respect to the imaginary time~\cite{Hele-JCP-2015} and the matrix elements are given explicitly in Sec.~I of the Supplement.
For notational convenience later on, the normal modes have the index $r$ that runs over $r=-(P-1)/2,\ldots,0,\ldots,(P-1)/2$; in contrast, the bead indices take the values $i=0,\ldots,P-1$, see \Fig{fig:path_NM_MM}.
Since the formulas are slightly different depending on the parity of $P$, it is chosen to be an odd number to get more symmetric expressions.
Importantly, taking the limit $P\to \infty$, which makes the Trotter factorization exact, is problematic due to the appearance of diverging terms.
These divergences can be, however, healed by an additional approximation that is described in the following.

The {\it second} approximation is to restrict the normal modes to the so-called Matsubara modes, i.e.\ to discard all higher normal modes with $|r|>\bar{M}$, where $\bar{M}:=(M-1)/2$ with $M \ll P$, see \Fig{fig:path_NM_MM} for a sketch.
Subsequently, the dependency of all physical quantities, in particular, the potential, the transition dipole moment as well as the time-dependent part on these higher ``non-Matsubara'' modes is neglected.
Practically, it implies that only smooth imaginary-time paths significantly contribute to the \TCF\ for all times $t$.
Under this assumption, the integrals corresponding to the non-Matsubara modes in \Eq{eq:phase-space} possess a Gaussian form and can be thus performed  analytically, see Sec.~II in the Supplement.
The resulting approximate time-independent part can be written down compactly as
\begin{align}
\label{eq:A-term_Mats}
\t{A}_l(\Q,\P) & \approx \frac{ \bar{M}!^2}{2 \pi \hbar^{M}Z} {D}^{g}_f (x_l(\Q)) \e^{ -\beta \left [ \t{H}_l(\Q,\P) + \im \T{\P} \MW \Q \right ] }  \enskip ,
\end{align} 
where the anti-symmetric anti-diagonal matrix $(\MW)_{r,s}:=\w_r \delta_{r,-s} $ contains the well-known Matsubara frequencies $\w_r:=2 \pi r / \beta \hbar$.
For the sake of brevity, $\Q$ and $\P$ represent the Matsubara modes only and $\t{A}_l(\Q,\P) := P^P {A}_l(\x(\Q),\p(\P))$ with $P^P$ stemming from the substitution of the Cartesian volume element by the normal mode one.
Note that we will define $\t{X}(\Q,\P):=X(\x(\Q),\p(\P))$ {\it without} the $P^P$ factor for any other arbitrary phase-space function $X$ in the following.
The classical Hamilton function in \Eq{eq:A-term_Mats}
\begin{align}
\label{eq:H_l}
\t{H}_l(\Q,\P):=\frac{1}{2 m} \T{\P} \P +  \t{U}_l(\Q)
\end{align}
is determined by the effective ring polymer potential for multiple \PESs\cite{Karsten_JCP_2018}
\begin{align}
\label{eq:effective_pot}
U_l(\x)& := \frac{1}{P} \sum_{i=0}^{l} \kappa_i V_f(x_i) +  \frac{1}{P} \sum_{i=l}^{P} \kappa_i V_g(x_i) \enskip ,
\end{align}
where $\kappa_i$ is equal to $1/2$ if $i$ corresponds to the first or the last summand, to $0$ if there is only one summand, which is the case if $l=0,P$, and to $1$ in all other cases.
%
%TODO: Check wether the section is correct.
Note that the harmonic spring term that is usually present in imaginary-time \PI\ methods has vanished upon applying the Matsubara approximation, see Sec.~II in the Supplement.
Note further that in the case $\l \in \{ 0,\beta \}$, with $\{ \bullet \}$ representing a set, the effective potential is exclusively assembled from the initial and final state potentials, respectively.
Having \Eq{eq:effective_pot} at hand, it becomes clear that the effective potential remains discontinuous at the points $x_0$ and $x_l$ in the limit $P \to \infty$ for $\l \in \, ]0,\beta[ \,$, where $ ]\bullet,\bullet[ $ refers to an open interval.
The question arises whether this is in the spirit of the Matsubara approximation which permits only smooth and continuous imaginary-time paths, see \Sec{sec:Results} for a discussion.

Importantly, after integrating out the non-Matsubara modes, the time-independent part does not feature any divergence as $P \to \infty$.
This is a direct consequence of the Matsubara approximation and will be particularly relevant for the consideration of the time-dependent part in \Sec{sec:time-dep-part}.
Interestingly, setting $V_g(x)=V_f(x)$, i.e.\ considering a single-\PES\ problem, and integrating \Eq{eq:shifted_TCF} over $\l$ from $0$ to $\beta$ yielding the Kubo transform, leads to the expression derived in Ref.~\citenum{Hele-JCP-2015}, even though the local harmonic approximation to the \PES\ and the linear approximation to the dipole, \Eqs{\ref{eq:V_harm},\ref{eq:dip_lin}}, were not utilized therein.

\subsection{The time-dependent part}
\label{sec:time-dep-part}

In order to evaluate the time-dependent part, we first consider its time derivative.
Since the factors in \Eq{eq:time-dep-part} are uncoupled, it is sufficient to consider them individually.
The time derivative of the $i$-th factor reads
\begin{align}
\label{eq:derivative-factor}
\dot{B}^i_l(x_i,p_i,t) & =  \frac{\im}{\hbar} \int \diff \Delta_i \ \e^{-\im {p}_i\Delta_i/\hbar}  \mBra{x_i^{+}}[\mOp{\H}_i \mOp{\mathcal{O}}_i(t)-\mOp{\mathcal{O}}_i(t)\mOp{\H}'_i] \mKet{x_i^{-}} \enskip,
\end{align}
where $\mOp{\mathcal{O}}_i(t):= \exp[\im \mOp{\H}_i t/\hbar ]\mOp{\mathcal{O}}_i \exp[-\im \mOp{\H}'_i t/\hbar ]$ with $\mOp{\mathcal{O}}_i$ defined in \Sec{sec:phase_space}.
Recognizing the Wigner transform in \Eq{eq:derivative-factor}, the time derivative can be {\it exactly} rewritten in terms of a quantum Liouvillian, see e.g.\ Ref.~\citenum{Shemetulskis-JCP-1992}, as $\dot{B}^i_l(x_i,p_i,t)=\mOp{\mathcal{L}}^i_l(x_i,p_i)B^i_l(x_i,p_i,t)$, where
\begin{align}
\mOp{\mathcal{L}}^i_l(x_i,p_i) & := \frac{p_i}{m} \frac{{\partial}}{\partial x_i} - \frac{1}{\hbar} [\mathcal{V}_i(x_i)+\mathcal{V}'_i(x_i)] \sin\left(\frac{\hbar}{2} \frac{\overleftarrow{\partial}}{\partial x_i} \frac{\overrightarrow{\partial}}{\partial p_i} \right )+ \frac{\im}{\hbar} [\mathcal{V}_i(x_i)-\mathcal{V}'_i(x_i)] \cos\left(\frac{\hbar}{2} \frac{\overleftarrow{\partial}}{\partial x_i} \frac{\overrightarrow{\partial}}{\partial p_i} \right ) 
\end{align}
and the arrows on top of the partial derivatives imply the direction of their action.
Importantly, all terms $\mathcal{V}_i(x_i)-\mathcal{V}'_i(x_i)$ vanish apart from the case $i=0$, where $\mathcal{V}_0(x_0)-\mathcal{V}'_0(x_0)=V_f(x_0)-V_g(x_0)$, see definitions of $\mOp{\H}_i$ below \Eq{eq:time-dep-part}.
After summing up the individual parts and noting that mixed derivatives with respect to different indices $i$ vanish, one gets the derivative of the complete time-dependent part  as $\dot{B}_l(\x,\p,t)=\mOp{\mathcal{L}}_l(\x,\p)B_l(\x,\p,t)$, with the full and still exact quantum Liouvillian
\begin{align}
\label{eq:QuantumL}
\mOp{\mathcal{L}}_l(\x,\p) & :=  \frac{1}{m}\T{\p}{\grad}_{\x} - \frac{2 P}{\hbar}\U_l(\x)\sin\left( \frac{\hbar}{2} \T{\overleftarrow{\grad}_{\x}}\overrightarrow{\grad}_{\p}  \right ) + \frac{\im}{\hbar} [V_f(x_0)-V_g(x_0)] \cos\left(\frac{\hbar}{2} \T{\overleftarrow{\grad}_{\x}}\overrightarrow{\grad}_{\p} \right ) \enskip,
\end{align}
where 
\begin{align}
\label{eq:dyn-pot}
\U_l(\x):=\frac{1}{2P} [V_f(x_0)+V_g(x_0)] + \frac{1}{P} \sum_{i=1}^{P-1} \mathcal{V}_i(x_i) \enskip.
\end{align}
Following the same line of reasoning as in the previous section, the normal mode transform is employed leading to $\T{\overleftarrow{\grad}_{\x}}\overrightarrow{\grad}_{\p} =  \T{\overleftarrow{\grad}_{\Q}}\overrightarrow{\grad}_{\P}/P$, where it is important to note the downscaling by $P$.
Restricting the normal modes to the Matsubara modes and bearing in mind that the limit $P \to \infty$ of the time-\textit{independent} part exists, we can consider this limit for the time-dependent part as well.
Keeping only the leading terms in the Taylor series of sine and cosine in \Eq{eq:QuantumL} yields
\begin{align}
& \lim_{P \to \infty} \mOp{\t{\mathcal{L}}}_l(\Q,\P)  =  \frac{1}{m}\T{\P} {\grad}_{\Q} - [ \T{\grad}_{\Q} \t{\U}_l(\Q) ] {\grad}_{\P}   + \frac{\im}{\hbar} [V_f(x_0(\Q))-V_g(x_0(\Q))] 
\enspace ,
\end{align}
which is a \textit{classical}-like Liouvillian containing a scalar imaginary inhomogeneity that depends on the energy gap between the initial and final electronic states.
Importantly, neither an additional approximation nor any $\hbar \to 0$ limit has been applied; note that the latter would be a non-trivial task since the time-independent part still depends on $\hbar$.
Having the particular form of the Liouvillian at hand, one can write down the time-dependent part explicitly as 
\begin{align}
\t{B}_l(\Q,\P,t) & = \e^{\mOp{\t{\mathcal{L}}}_l(\Q,\P) t }\t{B}_l(\Q,\P,0) ={D}^{f}_g(x_0(t)) \exp\left\{\frac{\im}{\hbar}\int_0^{t} [V_f(x_0(\tau))-V_g(x_0(\tau)) ]\diff \tau \right\}
\enskip,
\end{align}
where $x_0(t):=\e^{\Re\mOp{\t{\mathcal{L}}}_l(\Q,\P) t }x_0(\Q)$ with $\Re \mOp{\t{\mathcal{L}}}_l$ meaning the real part of $\mOp{\t{\mathcal{L}}}_l$.
Alternatively to the formulation with Liouville operators, one could now define a classical Hamilton function that generates the trajectory $x_0(t)$ with $x_0(0)=x_0(\Q)$ via Hamilton's \EOMs\ for $\Q$ and $\P$ as it is spelled out explicitly in the following section.
%

%-----------------------------------------------------------------------
\subsection{The Matsubara approximation to the imaginary-time shifted TCF}
\label{sec:bringing-together}
%-----------------------------------------------------------------------

Putting together the results presented in the last two sections leads to the Matsubara approximation to the imaginary-time shifted correlation function
\begin{align}
\label{eq:final}
\nonumber
 C_{\l}(t)  \approx & \frac{ \bar{M}!^2}{2 \pi \hbar^{M}Z} \int \diff \Q \int \diff \P\, \\
\times  &   \e^{ -\beta \left [ \t{H}_l(\Q,\P) + \im \T{\P} \MW \Q \right ] }  {D}^{g}_f (x_l(0)) {D}^{f}_g(x_0(t))
 \exp\left\{\frac{\im}{\hbar}\int_0^{t}[V_f(x_0(\tau))-V_g(x_0(\tau))] \diff \tau \right\}
\enskip,
\end{align}
where the dynamics follows from the \EOMs
\begin{align}
\label{eq:EOMs}
\dot{\Q}=\frac{\P}{m}, \quad \dot{\P}=-\grad_{\Q} \t{\U}_l(\Q) \enskip.
\end{align}
The last two equations constitute one of the main theoretical results of this manuscript and the starting point for simulating the vibronic spectra of complex systems (after a straightforward generalization to a many-dimensional many-body case).
However, a few concerns are yet unresolved.
First, it is not clear which particular form of the potential $\t{\U}_l(\Q)$ is reasonable.
As it is discussed in the next section, this freedom might be beneficial.
Second and a more severe one, is the presence of the imaginary part $\T{\P} \MW \Q $ in the exponent in \Eq{eq:final}, which is responsible for the infamous sign problem, leading to an unsufficient statistical convergence, see \Sec{sec:sign-problem}.
%

%-----------------------------------------------------------------------
\subsection{Choosing the dynamics}
\label{sec:choosing}
%-----------------------------------------------------------------------

As it has been stated in \Sec{sec:phase_space}, the Hermitian operators $\mOp{\H}_i$ that determine the dynamics are, in principle, completely arbitrary.
However, one can think of two choices aiming at a reasonable simulation protocol.

First, in an attempt to avoid problems such as zero-point energy leakage,~\cite{Habershon-JCP-2009} it would be desirable to keep the density stationary.
Thus, one can define the Hamiltonians
\begin{align}
	\label{eq:H_k}
	\mOp{\H}_i:=\begin{cases}
		\mOp{H}_f\enspace, & 1 \leq i < l\\
		\mOp{H}_{\mathrm{av}}:=(\mOp{H}_g+\mOp{H}_f)/2\enspace , & 0<l<P\,,\ l=i \\
		\mOp{H}_g\enspace, & l < i \leq P-1 \enskip,\\
	\end{cases}
\end{align}
such that, at first glance, the equality $U_l(\Q)=\U_l(\Q)$ (see \Eqs{\ref{eq:effective_pot},\ref{eq:dyn-pot}} for their definitions) would lead to favorable equilibrium dynamics, since the density and the \EOMs\ are determined by the same potential.
Unfortunately, for $\l \in \{ 0,\beta \}$, the equality $U_l(\Q)=\U_l(\Q)$ {\it cannot} be achieved, since $U_{0}$ and $U_\beta$ are exclusively assembled from the initial and the excited state potential, respectively, whereas $\U_{0}$ and $\U_{\beta}$ will \textit{always} feature the averaged potential for $x_0$, see \Eq{eq:dyn-pot}.
%
%TODO: Check the section.
Further, for $\l \in \, ]0,\beta[ \,$, the Matsubara phase ($\T{\P} \MW \Q$ in \Eq{eq:final}) varies in time and so does the density, as it is shown in the Supplement, Sec.~III.
In contrast, this time dependence vanishes in the continuous limit for the single \PES\ problem and, hence, the Matsubara density becomes strictly stationary as $P \to \infty$.
Eventually, it seems impossible to find any set of $\mOp{\H}_i$ for any $\l$ that would lead to truly stationary dynamics for a vibronic spectroscopy study.
Another possible objection against this choice of the potential is the presence of the discontinuity at $x_0$ and $x_l$ that contradicts the path smoothness assumed within the Matsubara approximation, see the comments to \Eq{eq:effective_pot}.
Despite these facts, this particular choice of $\mOp{\H}_i$ will be employed and referred to as the {\it equilibrium method} in the following.

Second, by setting all $\mOp{\H}_i:=\mOp{H}_{\mathrm{av}}$ the discontinuity is  avoided at the price of obviously non-stationary dynamics.
This choice, dubbed as the \textit{average method}, is more compatible with smooth paths assembled from the Matsubara modes.
Importantly, both the equilibrium and the average setups recover the usual Matsubara dynamics for the single-\PES\ problem, i.e.\ $V_g(x)=V_f(x)$.
Moreover, in contrast to known classical(-like) approaches to vibronic spectra such as the \DCL, the \ACL~\cite{Shemetulskis-JCP-1992} and the method presented in Ref.~\citenum{Karsten_JCP_2018}, both choices of the dynamics recover the exact imaginary-time shifted \TCF\ of a harmonic oscillator model, as it is shown in \Sec{sec:Results}.

Here, we have considered only the two aforementioned logically sound choices for the potentials.
Nonetheless, the presented formalism offers in principle infinitely many possible approaches to vibronic spectroscopy that might or might not lead to efficient simulation protocols.

%-----------------------------------------------------------------------
\subsection{Circumventing the sign problem}
\label{sec:sign-problem}
%-----------------------------------------------------------------------

\subsubsection{The conventional ansatz}

As it has been already stated, the Matsubara approximation to the quantum \TCF\ suffers from the sign problem due to the presence of the imaginary part $\T{\P} \MW \Q $ in the exponent in \Eq{eq:final}.
However, this imaginary part can be removed by transforming the Matsubara momentum to the complex plane, independently on the form of the physical potential~\cite{Hele-JCP-2015}
\begin{align}
\P \mapsto \P+\im m \MW \Q \enskip.
\end{align}
It is straightforward to show that the exponent then becomes
\begin{align}
\label{eq:RPMD1}
\t{H}_l(\Q,\P)+\im \T{\P} \MW \Q  \mapsto \t{H}_l(\Q,\P)+\frac{1}{2} m \T{\Q} \T{\MW} \MW \Q 
\enskip ,
\end{align}
which is purely real and thus there is no sign problem occurring.
Unfortunately, the new Liouvillian
\begin{align}
\label{eq:RPMD2}
\Re\mOp{\t{\mathcal{L}}}_l(\Q,\P) \mapsto \Re\mOp{\t{\mathcal{L}}}_l(\Q,\P)-m  \T{\Q}\T{\MW} \MW \nabla_{\P} + \im  [\T{\P} \MW  \nabla_{\P}-\T{\Q} \MW  \nabla_{\Q}]
\end{align}
would generate unstable complex trajectories yielding similar statistical convergence issues as the sign problem itself.

One common ansatz to ultimately avoid the sign problem is to simply discard the imaginary part of the Liouvillian.~\cite{Hele-JCP-2015}
Applying this approximation to \Eq{eq:RPMD2} leads to a \RPMD-like method for multiple \PESs, since the additional terms on the right hand sides of \Eqs{\ref{eq:RPMD1},\ref{eq:RPMD2}} correspond to the well-known spring terms.~\cite{Tuckerman-Book}
Importantly, if one picks the equilibrium choice for the dynamics and considers only $\l \in \, ]0,\beta[$, then the resulting approximation to $C_\l(t)$ coincides with the one presented in Ref.~\citenum{Karsten_JCP_2018} in the limit $M,P \to \infty$.
Hence, the method carried out therein can be viewed as an approximation to the Matsubra dynamics presented here.
Nevertheless, this relation does not hold at the imaginary-time borders, i.e.\ $\l \in \{ 0,\beta \}$ since the bead located at $x_0$ experiences the averaged potential, see \Eq{eq:dyn-pot}, which is not present in Ref.~\citenum{Karsten_JCP_2018}.
We note in passing that other common approximations circumventing the sign problem, such as  \TRPMD~\cite{Rossi-JCP-2014} and \CMD,~\cite{Cao-JCP-1993} would be equally transferable to the multiple \PESs\ case, starting from the developed Matsubara dynamics ansatz.

\subsubsection{The modified method}

Utilizing the harmonic model system introduced in \Sec{sec:Comp_det} allows one to write down explicit expressions for various \TCFs.
In particular, a comparison of the \RPMD\ expression for the standard \TCF, $C_0(t)$, to its Matsubara counterpart suggests an interesting though \textit{ad hoc} modification of the simulation protocol, see Sec.~IV of the Supplement.
Substituting the exponent of the complex density from \Eq{eq:final} as
\begin{align}
\label{eq:modified}
  \frac{1}{2m} \T{\P}\P + \frac{1}{2}m\w^2 \T{\Q}\Q  +\im \T{\P} \MW \Q   \mapsto   \frac{1}{2m} \T{\P} ( 1+{\T{\MW}\MW}/{\w^2} ) \P + \frac{1}{2}m \w^2\T{\Q}  ( 1+{\T{\MW}\MW/\w^2} ) \Q  
\end{align}
and keeping the \EOMs\ given in \Eq{eq:EOMs} unchanged, recovers the analytical Matsubara expression for the harmonic oscillator model considered.
Consequently, this \textit{modified method} recovers the exact \TCF\ in the $M,P \to \infty$ limit as the Matsubara method does.
However, since the resulting modified density is real and non-negative, there is no sign problem and thus the suggested modification leads to a strikingly more efficient protocol than the Matsubara approximation, see \Sec{sec:Results}.
Interestingly, in contrast to the \RPMD-like protocol, this modified method features dynamics \textit{without} springs, which can cause artificial resonances with physical modes that may ruin the correct description of dynamical properties, as is well-known in vibrational spectroscopy.~\cite{Witt-JCP-2009,Habershon-JCP-2008}
Instead, there is an additional 
%SK: For me dynamics corresponds to the real-time evolution but I'm fine.
%SI: I reflected it now. 
spring term, which connects the momenta of the beads in the density but does not affect the dynamics.

Although being constructed in an \textit{ad-hoc} fashion, the modified density is directly related to the thermal Wigner function for the harmonic oscillator case, as it is shown in \Sec{sec:harmonic_num}.
The proposed modification is based exclusively on the expressions for $\l=0$ and, thus it expectedly fails for any case, where $\l \in\, ]0,\beta[$, see Sec.~V in the Supplement.
Inventing analogous modifications for these cases is non-obvious and, therefore, a rigorous justification of the replacement would be desirable to identify the missing ingredients.

In order to go beyond purely harmonic systems, the modified method has to be generalized to an anharmonic system.  
We suggest to perform the following modification
\begin{align}
\label{eq:modified_gen}
\frac{1}{2m} \T{\P}\P + \t{U}_0(\Q)  -\im  \T{\P} \MW \Q \mapsto   \frac{1}{2m} \T{\P} ( 1+{\T{\MW}\mMat{Y}^{-1}(\Q)\MW} ) \P + \t{U}_0(\Q) + \frac{1}{2}m\T{\Q}  {\T{\MW}\MW} \Q 
\enskip ,
\end{align}
while keeping again the original \EOMs\ from \Eq{eq:EOMs}.
The key quantity is the position-dependent matrix $\mMat{Y}(\Q)$, which maps the curvature of the ground-state potential along the imaginary-time path onto the Matsubara modes, i.e.
\begin{align}
\mMat{Y}(\Q):= \frac{1}{m} \MT \left ( \frac{\partial^2 V_g }{\partial x_i^2 } \delta_{ij} \right ) \T{\MT} \enskip .
\end{align}
This seems to us as the most flexible generalization to anharmonic systems, while keeping the structure as close as possible to that in \Eq{eq:modified} and having it as the limiting case if $V_g(x)$ is perfectly quadratic.
As a word of caution, it should be noted that the inverse $\mMat{Y}^{-1}$ becomes ill-defined at the turning points of the ground-state \PES\ and that the modified density can become unbound if $V_g(x)$ possesses regions with a negative curvature.
However, for the present study these problems are avoided owing to the particular form of the quartic potential as given in \Sec{sec:Comp_det}.
It remains to be seen whether this new but heuristic simulation protocol can be viewed as a systematic approximation to the orignal Matsubara method or if there is any relation to the Feynman-Kleinert Quasi-Classical Wigner method.~\cite{Smith_JCP_2015}
Note that in case of $\l=0$ all aforementioned methods tend to the well-known \ACL\ one~\cite{Shemetulskis-JCP-1992,Rabani-JCP-1998,Egorov-JCP-1998} as $P \to 1$.
This is in contrast to Ref.~\citenum{Karsten_JCP_2018}, where the $\l=0$ and $P=1$ case coincides with the \DCL\ method.

%-----------------------------------------------------------------------
\section{Computational details}
\label{sec:Comp_det}
%-----------------------------------------------------------------------

In order to analyze the methodology presented in \Sec{sec:Theory}, we consider in \Sec{sec:harmonic} a rather simple but still non-trivial model system consisting of two shifted harmonic \PESs, $V_a(x)=m\omega^2 (x-x_a)^2 /2$ with $a=g,f$ and the same harmonic frequency $\w$.
Consequently the energy gap $V_f(x)-V_g(x)$ is a linear function of the coordinate.
The exact \QM\ imaginary-time shifted \TCF\ for such a system in the Condon approximation, i.e.\ $D_g^f(x)\approx 1$, is known and is compared to its approximations in \Sec{sec:harmonic_ana}.

For practical simulations performed in \Sec{sec:harmonic_num}, the parameters for the OH diatomic adopted from the qSPC/Fw model~\cite{Paesani2006} are employed, that is $m=m_{\mathrm{OH}}=1728.0\,$au, $\w=\w_{\mathrm{OH}}=0.0177\,$au, $x_g=0$ and $x_f=0.5\,$au.

In order to investigate a more realistic scenario in \Sec{sec:Morse}, we have modeled the OH-diatomic by two displaced anharmonic oscillators, where
\begin{align}
\label{eq:Morse}
V_a(x)=E_a\left [ \alpha^2_a(x-x_a)^2 - \alpha^3_a(x-x_a)^3 +\frac{7}{12}\alpha^4_a(x-x_a)^4 \right ]
\end{align}
is a quartic expansion of the Morse potential.
The parameters for the electronic ground state, $E_g=0.185\,$au and $\alpha_g=1.21\,$au, are again adopted from the qSPC/Fw model for water~\cite{Paesani2006} with $x_g=0$.
Similarly to the choice made in Ref.~\citenum{Karsten_JCP_2018}, the excited state differs from the ground state by a displacement of $x_f-x_g=0.22\,$au and a lower stiffness of $\alpha_f=0.86\alpha_g$, whereas the dissociation energies $E_f=E_g$ are the same. 
The equilibration as well as the sampling have been performed at 300\,K using a standard Langevin thermostat~\cite{Bussi-PRE-2007} and the \EOMs\ have been integrated employing the velocity-Verlet algorithm.
A reasonably large number of 10000 microcanonical trajectories starting from uncorrelated initial conditions has been used for all the methods and systems.
Only the Matsubara method with $P=9$ has required 10 times more trajectories to converge, see below.
The quantum results have been obtained in the harmonic basis of the electronic ground state potential using 50 eigenstates.
In order to avoid artifacts of the finite Fourier transform, the spectra have been convoluted with a Gaussian window $\exp[-\w^2/2\sigma^2]$, where $\sigma=0.002\,$au.

Although setting $M=P$ contradicts the Matsubara approximation, where $M \ll P$ is implied, see \Sec{sec:time-ind-part}, this choice still yields reasonable results for various simulation protocols, as it will be discussed in \Sec{sec:Results}.
In such a scenario, the correlation function originates from purely classical dynamics of $P$ \textit{independent} particles that are only connected via the Matsubara phase in the density, \Eq{eq:final}.
In the following discussion, we will therefore make the distinction between this dynamics of independent classical particles, i.e.\ $M=P$, and the {\it true} Matsubara dynamics, involving only smooth imaginary-time paths, i.e.\ $M<P$.
Note that for the \RPMD-like method the beads are coupled via the spring term in any case.

%-----------------------------------------------------------------------
\section{Results and discussion}
\label{sec:Results}
%-----------------------------------------------------------------------

%-----------------------------------------------------------------------
\subsection{Analysis for the harmonic oscillator model}
\label{sec:harmonic}
%-----------------------------------------------------------------------

For the harmonic system introduced in \Sec{sec:Comp_det}, the exact \QM\ imaginary-time shifted \TCF, the corresponding Matsubara approximation as well as the expression resulting from the \RPMD-like ansatz and the modified method can be written down explicitly as functions of $M$ and $P$.
%
%TODO: Check the section number.
These analytical expressions, written down in Sec.~IV of the Supplement, serve as a basis for the following discussion that suggests several favorable simulation scenarios, subsequently employed in \Sec{sec:harmonic_num} and \Sec{sec:Morse}.

\subsubsection{Analytical results}
\label{sec:harmonic_ana}

The developed mathematical expressions are utilized for an applicability analysis of the considered methods, with respect to the relation between $M$ and $P$ and a choice of $\l$, see Table.~\ref{tab:applicability}.
The criterion for the applicability is the correct limit of the \TCF\ as $M,P \to \infty$ for the aforementioned harmonic oscillator system.
To reiterate, common classical(-like) approaches to vibronic spectroscopy, such as the \DCL\ and the \ACL\ do not yield the correct \TCF\ and thus spectrum with any parameter setup.~\cite{Karsten_JCP_2018}

\begin{table}[H]
\begin{center}
\begin{tabular}{|c||c|c|c|c||c|c|c|c||c|c|}
\cline{2-11} 
\hline 
\multicolumn{1}{|c||}{approximation} & \multicolumn{4}{c||}{Matsubara} & \multicolumn{4}{c||}{RPMD-like} & \multicolumn{2}{c|}{modified}\tabularnewline
\hline 
EOMs & \multicolumn{2}{c|}{equilibrium} & \multicolumn{2}{c||}{average} & \multicolumn{2}{c|}{equilibrium} & \multicolumn{2}{c||}{average} & equilibrium & average\tabularnewline
\hline 
$\lambda \in$ & $\{0,\beta\}$ & $]0,\beta[$ & $\{0,\beta\}$ & $]0,\beta[$ & $\{0,\beta\}$ & $]0,\beta[$ & $\{0,\beta\}$ & $]0,\beta[$ & $\{0\}$ & $\{0\}$\tabularnewline
\hline 
$M=P$ & ${\color{green}\boldsymbol{+}}$ & ${\color{green}\boldsymbol{+}}$ & ${\color{green}\boldsymbol{+}}$ & ${\color{green}\boldsymbol{+}}$ & $\boldsymbol{{\color{red}-}}$ & $\boldsymbol{{\color{red}-}}$ & $\boldsymbol{{\color{red}-}}$ & $\boldsymbol{{\color{red}-}}$ & ${\color{green}\boldsymbol{+}}$ & ${\color{green}\boldsymbol{+}}$\tabularnewline
\hline 
$M<P$ & $\boldsymbol{{\color{red}-}}$ & $\boldsymbol{{\color{red}-}}$ & ${\color{green}\boldsymbol{+}}$ & $\boldsymbol{{\color{red}-}}$ & $\boldsymbol{{\color{red}-}}$ & $\boldsymbol{{\color{red}-}}$ & $\boldsymbol{{\color{red}-}}$ & $\boldsymbol{{\color{red}-}}$ & $\boldsymbol{{\color{red}-}}$ & ${\color{green}\boldsymbol{+}}$\tabularnewline
\hline 
\end{tabular}
  \caption{\label{tab:applicability}
Applicability of the considered methods summarized in the first two rows to recover the exact result for the shifted harmonic oscillators in the limit $M,P\to \infty$.
Green plus signs mark the success, whereas red minus signs represent a failure.
In the third row, the curly braces indicate a set, while $]\bullet,\bullet[$ refers to an open interval.}
\end{center}
\end{table}

Let us start with the Matsubara dynamics, that is \Eqs{\ref{eq:final},\ref{eq:EOMs}}, carried out via average and equilibrium methods, as is discussed in \Sec{sec:choosing}.
Both methods yield the exact imaginary-time shifted \TCF\ for any $\l$ only for $M=P$, which contradicts the original idea of the Matsubara approximation.
However, since the equilibrium method is not exact for $M<P$ irrespectively of the value of $\l$,  $M=P$ remains the only reasonable setting for this choice of the \EOMs, see \Sec{sec:harmonic_num} and \ref{sec:Morse_Spec} for a discussion of its ability to approximate the quantum \TCF.
The only combination that still yields exact results with $M<P$ is the average method with $\l \in \{0,\beta\}$.
Solely for this choice, true Matsubara dynamics (with $M < P$) yields the exact quantum \TCF\ for the harmonic oscillator.
Importantly, for $\l \in\, ]0,\beta[$ none of the considered choices for the dynamics leads to the exact result if $M<P$.
This indicates that the Matsubara approximation, i.e.\ taking only smooth imaginary-time paths into account, is incompatible with with the discontinuity in the potential for the points $x_{0}$ and $x_l$, see \Sec{sec:time-ind-part} and \Sec{sec:choosing}.

\subsubsection{Numerical results}
\label{sec:harmonic_num}

In this section, the findings from \Sec{sec:harmonic_ana} are supplemented by a performance comparison between all the methods via numerical simulations at ambient temperature for a harmonic oscillator system, which mimics an isolated OH bond of a water molecule, see \Sec{sec:Comp_det}.
The focus is on on the convergence with respect to statistics and the number of beads and Matsubara modes, i.e.\ $P$ and $M$, respectively.

\begin{figure}
\includegraphics[width=0.99\columnwidth]{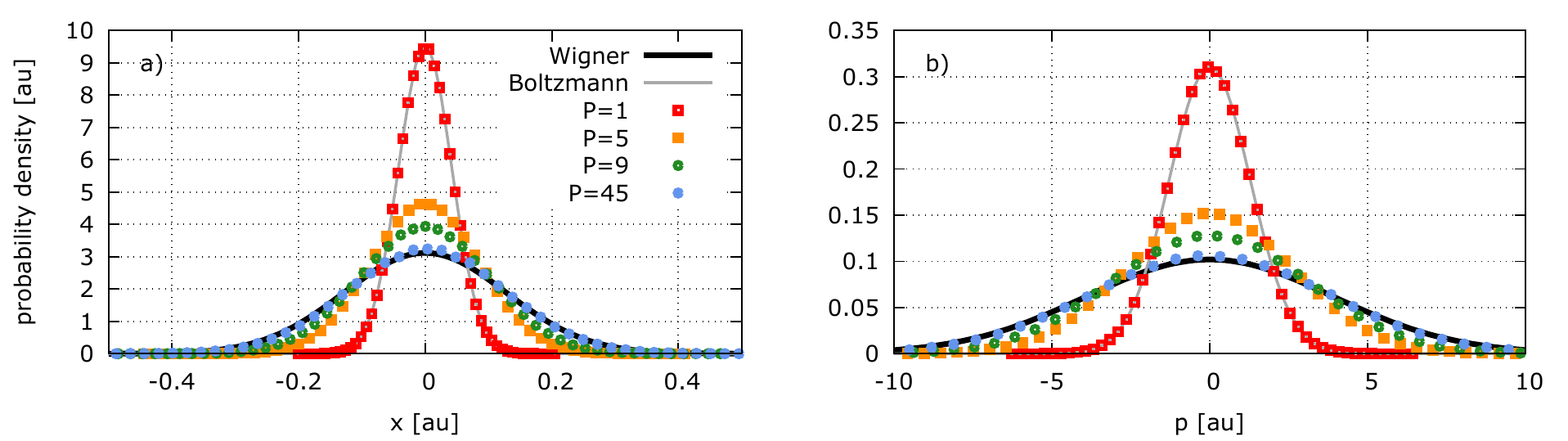}
\caption{\label{fig:Distribution} Histograms of the positions, panel a), and momenta, panel b), as sampled by the modified method for various values of $P$. The classical Boltzmann and the \QM\ Wigner functions are plotted as solid lines. Note that the full two-dimensional Wigner function can be constructed as the Cartesian product of both curves.}
\end{figure}

Let us first examine probability densities for the bead positions and momenta sampled by the modified method, \Eq{eq:modified}, see \Fig{fig:Distribution}.
Interestingly, they tend to the exact result given by the well-known thermal Wigner function of the harmonic oscillator, not only for positions, which is natural for \PI\ methods, but also for momenta.
Nonetheless, it is clear that such a correspondence cannot be general, as i) Wigner functions can become negative, which cannot be achieved by the suggested protocol; ii) the harmonic frequency $\w$ appearing on the r.h.s.\ of \Eq{eq:modified} cannot be unambiguously defined for a general system, see \Sec{sec:Morse} for suggestions.

\begin{figure}
\includegraphics[width=0.99\columnwidth]{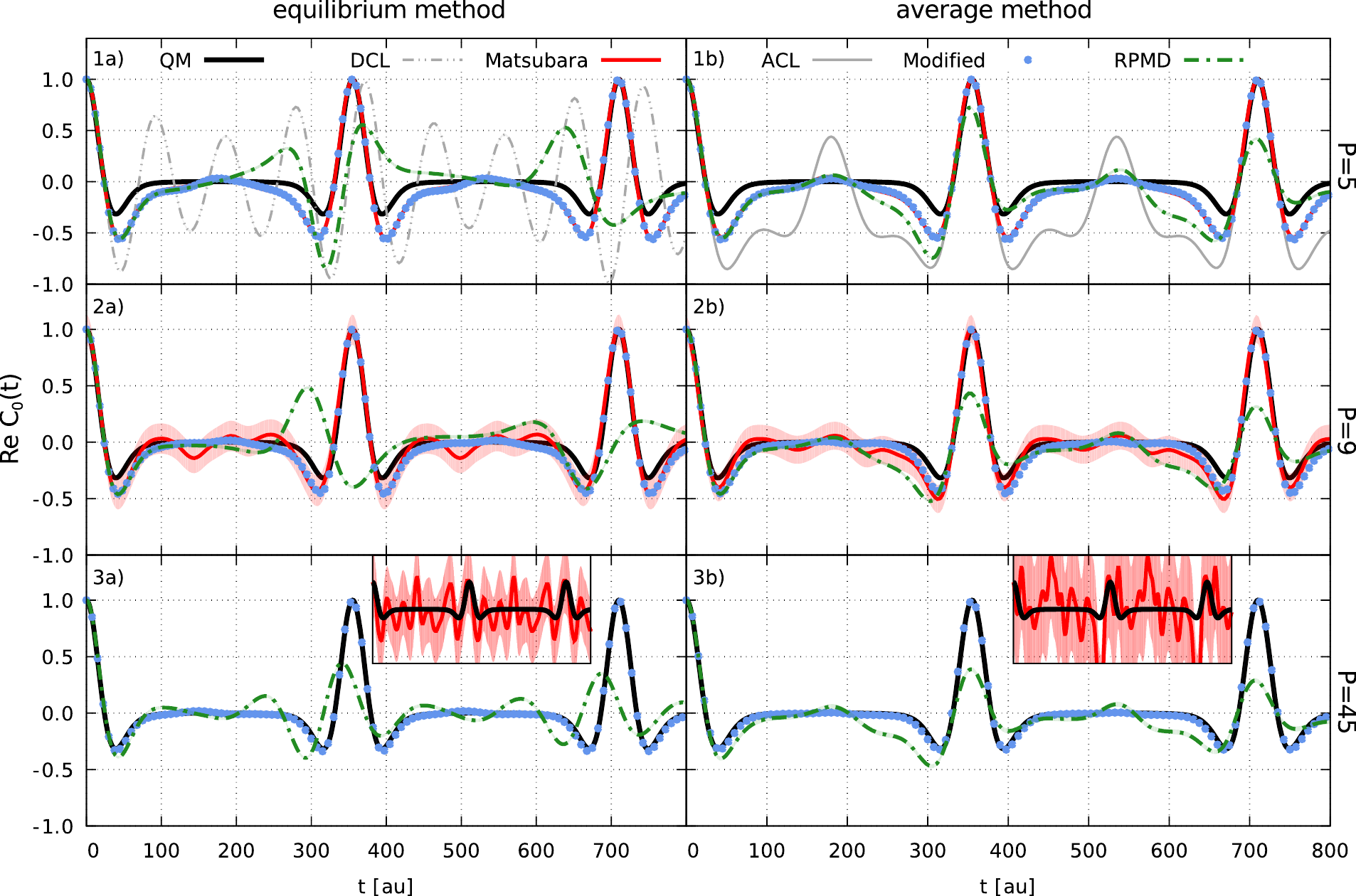}
\caption{\label{fig:Harmonic}
The real part of standard correlation functions, $C_0(t)$, for the harmonic system, see \protect\Sec{sec:Comp_det}.
The left column, panels a), exhibits the results of the equilibrium methods, whereas panels b) contain the results of the average methods for $M=P$.
The first row, panels 1), correspond to the case $M=P=5$, apart from the \DCL\ and \ACL\ curves which naturally imply $M=P=1$. 
The second row, panels 2), features $M=P=9$ and panels 3) show $M=P=45$ data, with insets exhibiting the results of the Matsubara method.
Transparent areas indicate the 95\% confidence intervals (which are not visible for most of the curves) and the lines and points represent the statistical mean.
}
\end{figure}

Switching to the \TCFs, we first restrict ourselves to $\l=0$ and $M=P$, see \Fig{fig:Harmonic}, where the left/right column corresponds to the equilibrium/average method.
For each method, the degree of statistical convergence is visualized by transparent areas representing the 95\,\% confidence interval around the mean values that are depicted as lines and points; note that for most of the methods these areas are within the thickness of the line.

\paragraph*{Independent classical dynamics.}
Starting from the common classical ($P=1$) approaches, one sees that both \DCL, \Fig{fig:Harmonic} panel 1a), and \ACL, panel~1b), fail to reproduce the correlation function, although the \ACL\ \TCF\ captures more features of the \QM\ curve.
For $P=5$, still panels 1) therein, one can see that the curves resulting from the Matsubara and the modified method coincide as expected and improve qualitatively over the \ACL\ \TCF, which is their common $P \to 1$ limit.
In case of the \RPMD\ ansatz, the average method performs much better than the equilibrium one, see panels 1a) and 1b), correspondingly.

When the number of beads is further increased to nine, panels 2) in \Fig{fig:Harmonic}, the first traces of the sign problem in the Matsubara method appear, as is manifested by red areas therein.
In contrast, the modified method still features an excellent statistical convergence and the curve gets closer to the exact \QM\ result.
Apart from the vicinity of the first minimum of the \TCF, both \RPMD-like methods do not improve, which underlines that \RPMD\ is a short-time approximation to the exact \QM\ dynamics.

Finally, taking a large number of beads ($P=45$), which is a typical value to reach convergence for an OH diatomic at 300\,K, as is confirmed in \Fig{fig:Distribution}, both versions of the Matsubara method exhibit severe problems with respect to statistical convergence, see insets in panels~3).
This illustrates the well-known fact that the Matsubara method is not directly applicable in practical numerical simulations.
In contrast, the modified method converges to the exact result without any issues and, importantly, without any additional costs with respect to the \RPMD-like method that in turn exhibits no further improvement with increasing $P$.

\paragraph*{True Matsubara dynamics.}

\begin{figure}
\includegraphics[width=0.99\columnwidth]{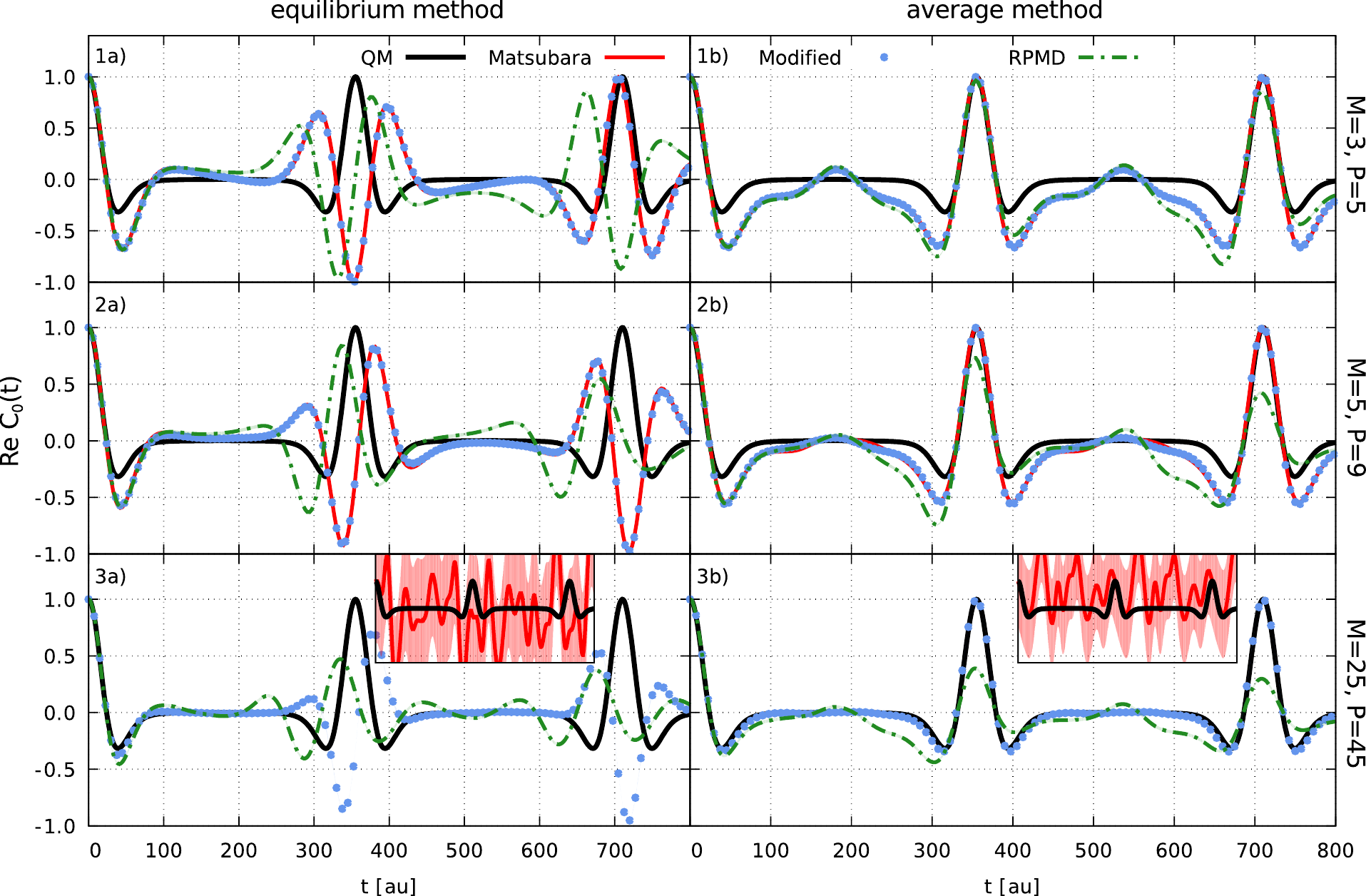}
\caption{\label{fig:Harmonic_M_lt_P}
Same scenario and color code as in \Fig{fig:Harmonic} but with the first row, panels 1), exhibiting the results for $M=3,P=5$. The second row, panels 2), features $M=5,P=9$ and panels 3) show $M=25,P=45$. }
\end{figure}

The Matsubara approximation implies that $M<P$, thus \TCFs\ with $M/P \approx 0.5$ for all the methods are shown in \Fig{fig:Harmonic_M_lt_P}, which has the same layout as \Fig{fig:Harmonic}.
The equilibrium Matsubara and the modified method (which coincide by construction) exhibit a phase shift with respect to the exact \QM\ curve for $M=3, P=5$, see panel 1a).
This bad performance is anticipated on the basis of the analysis presented in \Sec{sec:harmonic_ana}.
The \RPMD-like method fails in a similar way.
In contrast, the curves obtained by the average methods, panel 1b), are in phase with the \QM\ one.
Especially, the average \RPMD\ \TCF\ reveals much better agreement with the exact result than the equilibrium one.
Note that for the classical methods ($P=1$), the number of Matsubara modes cannot be smaller than the number of beads, thus \ACL\ and \DCL\ results are not shown. 

Turning to a larger number of modes and beads, $M=5, P=9$ in panels 2) of \Fig{fig:Harmonic_M_lt_P}, one notices immediately that in contrast to \Fig{fig:Harmonic} the Matsubara methods do not feature issues with the statistical convergence.
However, this is only due to the fact that the Matsubara density is assembled by 5 modes only, thus, the statistical behavior is effectively the same as for the case $M=P=5$.
Importantly, all the average versions approximate the \QM\ curve much better, which is especially true for the \RPMD-like method, compare panel 2a) and 2b).

Approaching the convergence with respect to the number of Matsubara modes and beads, i.e.\ for $M=25, P=45$, panels 3) in \Fig{fig:Harmonic_M_lt_P}, the Matsubara method suffers again from the sign problem, see insets.
The average version of the modified method converges to the exact result without any issues, however, as expected, the equilibrium one as well as the \RPMD\ method do not.

To conclude, the suggested modification strikingly outperforms the Matsubara method with respect to the statistical convergence, 
while yielding the same results for the case $\l=0$.
Moreover, the modified method surpasses the \RPMD-like ansatz with respect to the quality of the results.
As it has been anticipated from the mathematical analysis, the equilibrium versions of the Matsubara and the modified methods lead to reasonable results only if $M=P$.
The average methods outperform the equilibrium ones if $M < P$, which is especially true for the \RPMD\ ansatz.
At this point there is no pratical advantage of the quasi-classical dynamics ($M<P$) with respect to the independent classical dynamics ($M=P$).
However, since the considered system consists of two one-dimensional harmonic oscillators this is not surprising.
Whether such a statement can be transferred to a more realistic anharmonic case is discussed in \Sec{sec:Morse}.

%SK: TODO: To be checked in the final stage.
Finally, none of the presented methods can practically reproduce the exact imaginary-time shifted \TCF\ for the case $\l \in\, ]0,\beta[ \,$, see Sec.~V in the Supplement.
In particular, the equilibrium Matsubara method converges formally to the exact result if $M=P$ but is suffering from the sign problem.
The \RPMD-like method does not lead to an acceptable approximation to the quantum \TCF, as in all the other cases considered.
Unfortunately, the modified method was derived just for $\l=0$ and naive use of this modification for $\l>0$ expectedly fails.
Still, the expressions that have been derived in \Sec{sec:Theory} for a general value of $\l$ might serve as a basis for future work and finally may lead to a powerful simulation protocol in combination with the generalized \TCF\ formalism.~\cite{Karsten_JCP_2018}

\subsection{Anharmonic oscillators}
\label{sec:Morse}

\subsubsection{Generalization of the modified method}
\label{sec:modified_gen}

\begin{figure}
\includegraphics[width=0.99\columnwidth]{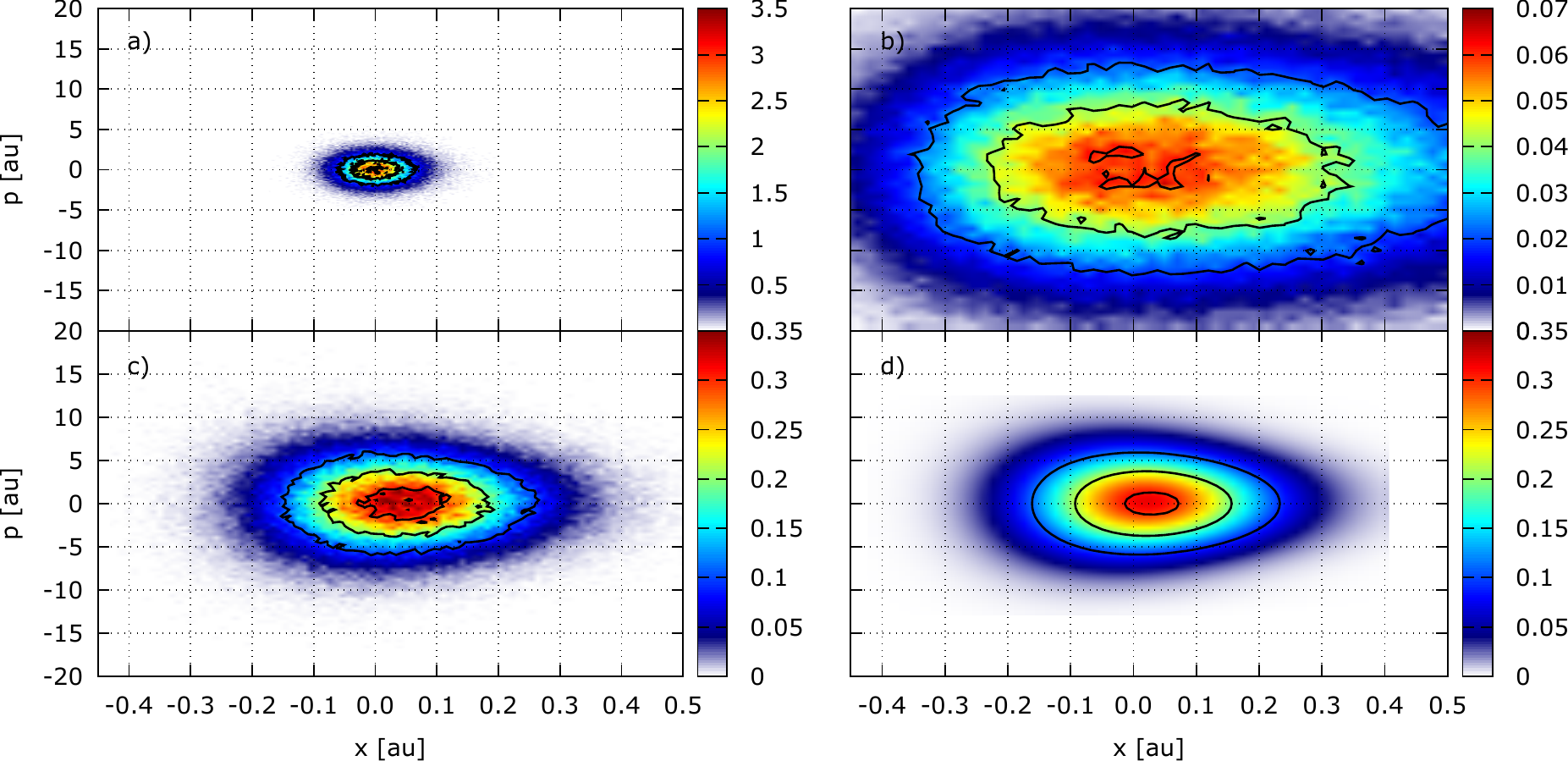}
\caption{\label{fig:Wigner_Morse} Phase-space probability densities for the anharmonic model system sampled by the considered methods.
Panel a) the classical Boltzmann distribution; panel b) the absolute value of the complex Matsubara density and panel c) the modified density, both with $P=45$; panel d) the exact Wigner function.
}
\end{figure}

Since the modified density tends to the exact Wigner function for the harmonic oscillator, see \Fig{fig:Distribution},  we investigate the ability of the generalized modified method to reproduce the exact Wigner function of the anharmonic system introduced in \Sec{sec:Comp_det}.
In \Fig{fig:Wigner_Morse}a) one can see the classical Boltzmann function for the electronic ground state, which is the common limit for all methods if $P=1$.
It is not surprising that the classical density is much more localized than the exact Wigner function seen in panel d); note the different scales for the color bars in the different panels.
In panel b), the absolute value of the complex density corresponding to the Matsubara method, \Eq{eq:final}, is shown for $P=45$.
This is effectively nothing else than the classical Boltzmann distribution but for a temperature that is $P$ times higher than in the classical case, panel a). 
For the present model this would correspond to a temperature of 13500\,K and, hence, the density covers a phase-space volume that is much larger than the classical one and significantly larger than the correct \QM\ one.
Only the cancellation due to the complex phase in the Matsubara density, removes the irrelevant contributions to the observables.
This \textit{indirect} sampling of the correct distribution is at the heart of the sign problem and is thus responsible for the poor convergence of the Matsubara method.
In contrast, the generalized method, \Eq{eq:modified_gen}, approximates the exact thermal Wigner function \textit{directly} with a remarkable accuracy, see panel c).
Importantly, it is even able to reproduce the ``egg shape'' of the correct Wigner function, which is an inherently quantum-statistical effect, since it requires the coupling of position and momenta in the density, which is not present in the classical case.

In resume, the suggested generalization captures important quantum statistics even for the anharmonic case, although
any negativity in the Wigner function remains outside reach and the numerical instabilities in the regions of a vanishing or negative curvature of the potential are still awaiting for a proper treatment.

\subsubsection{Vibronic absorption spectra for the anharmonic system}
\label{sec:Morse_Spec}

\begin{figure}
\includegraphics[width=0.99\columnwidth]{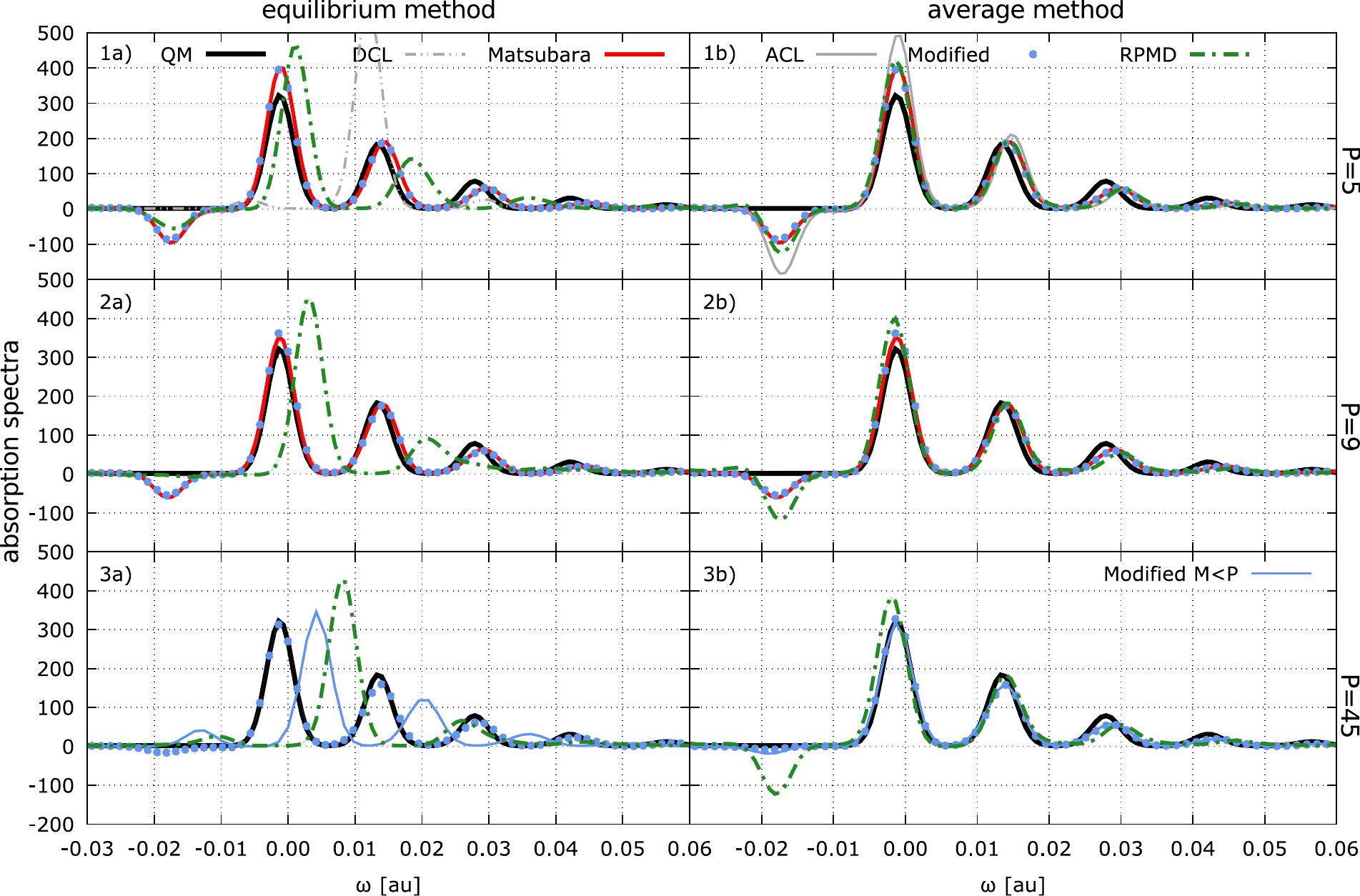}
\caption{\label{fig:Anharmonic} 
Absorption spectra for two displaced anharmonic oscillators, see \Eq{eq:Morse}. Same color code as in \Fig{fig:Harmonic} but with the thin blue lines in the lower row representing the setup $M=25,P=45$.}
\end{figure}

After discussing static properties of the methods, the dynamical observables, i.e.\ vibronic absorption spectra are considered for the anharmonic system.
With the parameters given in \Sec{sec:Comp_det}, simulations have been performed yielding the results depicted in \Fig{fig:Anharmonic}, which has the same structure as \Fig{fig:Harmonic} but now showing the Fourier transforms of $C_0(t)$.

Starting with the description of the \QM\ spectrum, one recognizes a typical Franck-Condon progression with a Huang-Rhys factor smaller than 0.5, meaning that the maximal intensity is located at the 0-0 transition.
Switching to the approximations in panel 1a), the common \DCL\ method neither yields the correct spectral shape nor the correct peak positions, as has been observed before.~\cite{Rabani-JCP-1998,Egorov-JCP-1998,Karsten_JCP_2018}
In contrast, the result obtained with the \ACL\ method, see panel 1b), is in much better agreement with the \QM\ one, though a significant negative intensity below the 0-0 transition is present.
This is a consequence of the non-stationary dynamics, possibly leading to all kinds of unphysical artifacts in spectra of more complex systems.~\cite{Karsten_JCP_2018}

When it comes to more modes and beads, i.e.\ $M=P=5$ in panels 1) of \Fig{fig:Anharmonic}, the modified method still coincides with the Matsubara method although this is generally not expected for an anharmonic case. 
Both methods cannot completely avoid a negative intensity below the 0-0 transition, but the amplitude of the artifact is smaller than the ones produced by the \ACL\ and \RPMD-like methods.
For the latter, the average method, see panel~1b), performs significantly better than its equilibrium counterpart depicted in panel~1a).

Increasing $M$ and $P$ to nine improves the quality of the modified and the Matsubara method, which are still surprisingly similar, as it can be seen in panels 2) of \Fig{fig:Anharmonic}.
However, the Matsubara method has required ten times more trajectories than the modified one to yield statistically converged results.
Importantly, the artificial negativity is smaller by a factor of two than that for the case $M=P=5$.
The average version of the \RPMD-like approach, see panel 2b), improves further with respect to the peak intensities, whereas the negative intensity has not changed notably.
The opposite can be observed in panel 2a) for the equilibrium \RPMD-like method, where the negativity completely disappears, whereas the overall agreement with the \QM\ curve becomes worse.

It appears to be impossible to reach statistical convergence for the Matsubara method, if the number of beads and modes is increased to 45, even employing $10^6$ trajectories and, thus, no results are shown.
The average version of the \RPMD\ method does not improve significantly in this case and the quality of its equilibrium counterpart becomes even worse, see panels 3).
The modified method convergences again without any issues, while coming quite close to the exact \QM\ curve.
Importantly, the artificial negativity vanishes almost completely.
Finally, one can see that there is again no significant difference between the cases $M=P=45$ and $M=25,P=45$ when considering the modified average methods, see the solid line in panel 3b).
The equilibrium method fails again if $M<P$, as it can be seen in panel 3a).

To conclude, the Matsubara and the modified methods nearly coincide for all cases, where the Matsubara method statistically converges.
One can therefore expect that both would converge to very similar results in the limit $M,P \to \infty$.
In comparison to the more common methods, i.e.\ \DCL, \ACL\ and finally the \RPMD\ ansatz, the modified method yields much more accurate results, especially with respect to the negativities below the 0-0 transition.
%
%TODO: Put here the correct section of the Supplement in the very final stage.
These statements are additionally supported by the results for a similar anharmonic system possessing a larger displacement between the potentials, see Sec.~V in the Supplement.
Interestingly, even for such a notably anharmonic system, the truly quasi-classical dynamics with $M<P$ yields no benefit if compared to the methods with $M=P$.
However, this is by no means the ultimate conclusion and the impact of having continuous imaginary-time paths within the dynamics has to be investigated carefully, in particular using more complex systems featuring problems such as zero-point energy leakage.

%-----------------------------------------------------------------------
\section{Conclusions and outlook}
\label{sec:Conclusions}
%-----------------------------------------------------------------------

The Matsubara dynamics has been generalized to the multi-\PES\ case and thereby to vibronic spectroscopy.
It has been shown that the Matsubara dynamics can be derived from the imaginary-time shifted \TCF, assuming that the potential is locally harmonic and the transition dipole moment is locally linear.
The performed derivation left certain flexibility in choosing the Hamiltonians responsible for the real-time propagation.
Practically, this enables infinitely many possible simulation protocols, whereas we have employed two particular ones, termed equilibrium and average methods, which have physical foundation.
In order to circumvent the infamous sign problem plaguing the Matsubara-based methods, an \RPMD-like ansatz has been employed.
The comparison of the Matsubara expressions to the \QM\ \TCF\ for a system of two shifted harmonic oscillators has singled out cases that lead to numerically exact results in the limit of infinitely many Matsubara modes.
Note that common classical-like approaches to the vibronic spectrum, such as the \DCL\ and the \ACL, do not yield exact results for this system with any set of parameters.
Further comparison of the \RPMD\ expressions to the exact ones has suggested an \textit{ad hoc} but meaningful modification of the density and its generalization to the anharmonic regime has been developed.
In both scenarios, the harmonic and the anharmonic one, the sampled density reproduced the exact thermal Wigner function of the electronic ground state with remarkable accuracy.
Although it cannot reproduce the negativities of the Wigner function by construction, it still may be a reasonable and very convenient method to sample Wigner densities, which is a common task in quasi- and semiclassical methods.~\cite{Beutier-JCP-2014}
When it comes to approximating the absorption spectra, the modified method outperforms the Matsubara method with respect to the statistical convergence, while it surpasses the \RPMD-like ansatz with respect to the quality of the results.
Further, if true Matsubara dynamics with smooth imaginary-time paths is considered, all of the presented average methods lead to significantly better results than the equilibrium ones.
Although the advantage of quasi-classical dynamics over independent classical dynamics has not become apparent for the considered one-dimensional systems, the importance of taking only smooth paths into account has to be investigated carefully for more complex systems.
Unfortunately, for imaginary-time shifted \TCFs, $C_\l(t)$ with $\l>0$, none of the methods is directly applicable to realistic systems.
To improve on the last point, future work might be dedicated to find a more rigorous justification of the modified method and finally to combine the generalized \TCF\ formalism~\cite{Karsten_JCP_2018} with the methodology presented here.
Another interesting perspective could be the application of the suggested modified method to single-\PES\ studies such as \IR\ spectroscopy.
In contrast to the more common \RPMD\ and \CMD\ methods, one would expect that problems with spurious resonances or with the curvature of the ring polymer~\cite{Witt-JCP-2009} could be avoided by the modified ansatz.

\begin{acknowledgments}
	S.~K.\ acknowledges financial support by the Deutsche Forschungsgemeinschaft \mbox{(KU~952/10-1)}.
\end{acknowledgments}

\bibliography{./united}
\end{document}